\documentclass[prl,twocolumn,superscriptaddress,showpacs,preprintnumbers,floatfix]{revtex4}

\usepackage{graphicx} 

\begin{document}

\title{Detecting phonons and persistent currents in toroidal \\
        Bose-Einstein condensates by means of pattern formation}
\author{M.~Modugno}
\affiliation{CNR-INFM BEC Center, I-38050 Povo, Trento, Italy}
\affiliation{LENS and Dipartimento di Matematica Applicata, 
Universit\`a di Firenze,
via Nello Carrara 1, I-50019 Sesto Fiorentino, Italy}
\author{C.~Tozzo}
\affiliation{Scuola Normale Superiore, I-56126 Pisa, Italy}
\author{F.~Dalfovo}
\affiliation{CNR-INFM BEC Center, I-38050 Povo, Trento, Italy}
\affiliation{Dipartimento di Fisica, Universit\`a di Trento,
             I-38050 Povo, Italy}

\begin{abstract}
We theoretically investigate the dynamic properties of a
Bose-Einstein condensate in a toroidal trap. A periodic modulation
of the transverse confinement is shown to produce a density
pattern due to parametric amplification of phonon pairs. By
imaging the density distribution after free expansion one obtains
i) a precise determination of the Bogoliubov spectrum and ii) a
sensitive detection of quantized circulation in the torus.
The parametric amplification is also sensitive to thermal and
quantum fluctuations.
\end{abstract}

\date{\today}


\maketitle

Bose-Einstein condensates have recently been obtained with ultracold 
gases in a ring-shaped geometry \cite{rings1}. Other groups 
are proposing different techniques to get toroidal condensates 
\cite{rings2}. The aim is to create a system in which 
fundamental properties, like quantized circulation and persistent 
currents, matter-wave interference, propagation of sound waves and 
solitons, can be observed in a clean and controllable 
way. An important issue concerns the feasibility of 
high-sensitivity rotation sensors.

In this work we show that key properties of condensates in
toroidal traps can be measured by means of parametric amplification. 
This corresponds to an exponential growth of some excited modes of 
the system induced by a periodic modulation of an external parameter
\cite{landau}. We consider the modulation of the transverse
confinement, which is shown to drive a spatially periodic pattern 
in both density and velocity distributions as a consequence of 
the amplification of pairs of counter-rotating Bogoliubov phonons. 
If the trap is switched off, this pattern produces a peculiar flower-like 
density distribution of the freely expanding gas. The number of ``petals" 
and their shape provide a sensitive measure of the excitation spectrum 
and the superfluid rotation of the condensate.

We perform numerical simulations by integrating the
Gross-Pitaevskii equation \cite{rmp} for $N$ bosonic atoms of mass
$M$, confined in an external potential  $V_{\rm ext}$:
\begin{equation}
i\hbar\partial_t\psi=\left[-\frac{\hbar^2}{2M}\nabla^2 + V_{\rm
ext} +g|\psi|^2\right]\psi \; . \label{eq:GP}
\end{equation}
The mean-field coupling constant is given by $g=4\pi\hbar^2a/M$,
where the s-wave scattering length $a$ is assumed to be positive.
The order parameter of the condensate, $\psi({\bf r},t)$, is
normalized to $ \int d {\bf r} |\psi|^2 = N$ and may by written as
$\psi= n^{1/2} \exp(iS)$, where $n$ is the condensate density and
the phase $S$ is related to the superfluid velocity by ${\bf v}
=(\hbar/M) \nabla S$. We consider a condensate of
$N=2\times 10^5$ atoms of $^{87}$Rb confined in a torus of length
$L=2\pi R=100\,\mu$m by an axially symmetric potential which has 
a minimum, $V_{\rm ext}=0$, at $z=0$ and $r_\perp= R$. The trap is
harmonic and isotropic in the $(z,r_\perp)$-plane around this 
minimum, with frequency $\omega_\perp=2\pi\times1$ kHz.  With 
this choice the transverse width of the condensate, $r_0$, is 
significantly smaller than the  radius of the torus. This implies 
that curvature effects are almost negligible in the ground state 
and in the in-trap dynamics and one can replace the torus of 
radius $R$ with a cylinder of length $L$ with periodic boundary 
conditions \cite{note2D}. Curvature effects are instead important 
during the free expansion of the condensate, and therefore 
the full toroidal geometry is used for simulating the dynamics 
after the release from the trap. The choice made for the 
geometry and the parameters is intended to simulate feasible
experiments, but the effects we are going to show can be 
obtained with different types of traps and in a wide range
of parameters. 

The process that we want to study has three steps: i) the
condensate is initially prepared in the torus; ii) the transverse
harmonic potential is periodically modulated for a time $t_{\rm
mod}$; iii) both the trap and the modulation are switched off and
the gas freely expands for a time $t_{\rm exp}$.

We first solve the stationary GP equation and the Bogoliubov
equations obtained by linearizing Eq.~(\ref{eq:GP}) \cite{rmp} for 
a cylindrical condensate, by using the same numerical techniques 
as in \cite{bogoliubov}. We obtain the stationary 
solution $\psi_0({\bf r})$ and its excitation spectrum, namely  
the eigenfrequencies $\omega_{i}$ and quasiparticle amplitudes 
$u_i({\bf r})$ and $v_i({\bf r})$. The condensate at step (i) is 
assumed to be at equilibrium. Quantum and/or thermal 
fluctuations are included by means of the Wigner representation 
of bosonic fields \cite{wigner}. In practice, the function 
$\psi$, input of step (ii), is taken of the form $\psi = \psi_0 + 
\sum_{i} [c_{i} u_{i} + c_{i}^* v_{i}^*]$, where the sum extends 
over a wide set of Bogoliubov states including those 
which are expected to be relevant for the subsequent parametric 
amplification. The Wigner representation enters through the choice 
of the complex coefficients $c_{i}$, which are taken to have a 
randomly distributed phase, with constant probability in $[0, 2\pi)$,
and modulus, with Gaussian distribution of width $1/2 + [1 - 
\exp(\hbar\omega_{i}/k_BT)]^{-1}$. 

To simulate the in-trap dynamics at step (ii) we numerically
integrate Eq.~(\ref{eq:GP}) \cite{alg}. The external modulation 
is included by multiplying  $V_{\rm ext}$ by the factor
$f(t)=[1+s(t)A\cos(\Omega t)]$, where $A$ is the amplitude of the
modulation and $s(t)$ is an appropriate switching function. A
collective motion which could be easily excited by this perturbation 
is the radial breathing mode of frequency $2\omega_\perp$ 
\cite{breathing}. However, the processes that are of interest here 
are those involving longitudinal phonons at frequency lower than 
$2 \omega_\perp$. We also want that the transverse width and 
the density of the condensate oscillate in-phase with the external 
modulation. To this aim, the possible coupling with the  breathing 
mode has to be suppressed. This can be obtained by choosing 
$s(t)$ to be sufficiently adiabatic. 
In the following we will use an amplitude $A=0.1$, for which we  
find that a switching time of about $20$ ms is adequate in a wide 
range of $\Omega$. The entire modulation time, $t_{\rm mod}$, is 
instead chosen to be longer than $100$ ms. 

\begin{figure}
\centerline{\includegraphics[width=\columnwidth,clip=]{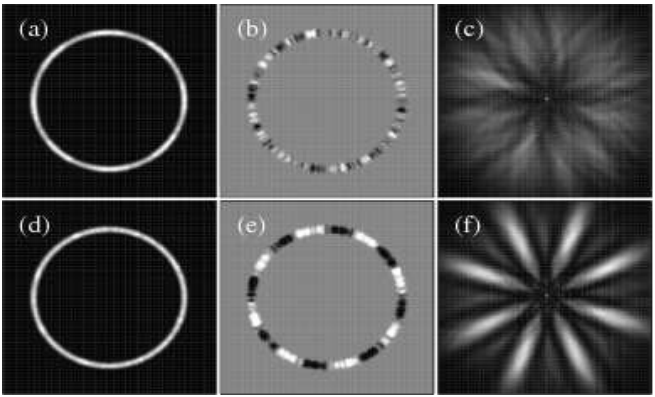}}
\caption{a) Density of a condensate in the toroidal 
trap after a transverse modulation with $\Omega=0.6~\omega_\perp$, 
$A=0.1$, and $t_{\rm mod}=130$ ms, corresponding to 
a maximum of visibility of the density pattern; the density 
exhibits $8$ oscillations and the relative density variation is
about $20$\%.  b) Tangential current for the same 
condensate. c) Density after $7$ ms of free expansion.  
d-e-f) Same as above but for $t_{\rm mod}=130.8$ ms, when 
the pattern visibility is maximum in the current distribution. 
Each box is $45~\mu$m per side. Densities and currents are integrated 
along the axis perpendicular to the torus. White (black) means
high (zero) density, while in frames (b) and (e) white and
black correspond to currents in opposite directions. 
} 
\label{fig:exp-dens}
\end{figure}

The transverse modulation is found to induce a longitudinal 
pattern formation \cite{cross} corresponding to a standing wave 
along the torus. An example is shown in Fig.~\ref{fig:exp-dens}(a) 
where the in-trap density distribution is plotted for a modulation 
of frequency $\Omega=0.6 \omega_\perp$ and duration 
$t_{\rm mod}=130$ ms. In this case the density oscillations 
along the torus have 8 maxima and correspond to a relative density 
variation of about $20$\%. This pattern is found to occur via the 
parametric instability of Bogoliubov phonons, with wavevector $k$, 
propagating along the torus in opposite directions. The mechanism 
is basically the same already described in \cite{kraemer,tozzo} 
for elongated condensates in a modulated 1D optical lattice and
it is analogous to the spontaneous pattern formation discussed 
in \cite{staliunas} in different geometries. It has also  
interesting similarities with the phenomenon of Faraday's 
instability \cite{cross,faraday} for classical fluids in annular 
resonators \cite{keolian}. 

The spatial periodicity of the standing wave in the torus is fixed 
by the wavelength of the most rapidly growing modes, determined 
by the Bogoliubov dispersion relation $\omega(k)$ and the boundary 
conditions. The parametric amplification spontaneously selects 
modes with frequency close to the resonance condition 
$\omega(k)=\Omega/2$ and wave vector $k=2\pi m/L$, where the 
integer $m$ represents the azimuthal angular momentum of the 
excitation and $2m$ is the number of nodes in the density pattern. 
The position of the nodes changes randomly at each realization,
being related to the phase of the initial fluctuations which are 
parametrically amplified. 

\begin{figure}
\centerline{\includegraphics[height=\columnwidth,angle=-90]{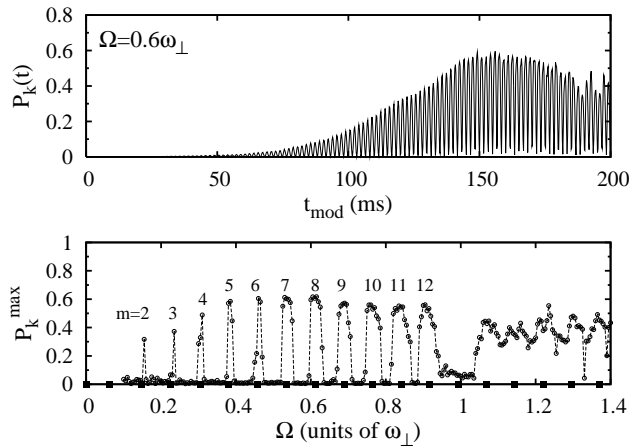}}
\caption{Top: Amplitude of the $\pm k$ components of the Fourier
transform of the order parameter along the torus, where the 
wavevector $k$ satisfies the resonance condition $\omega(k)=\Omega/2$. 
The amplitude is plotted as a function of $t_{\rm mod}$ for 
$\Omega=0.6 \omega_\perp$ and $A=0.1$.  Bottom: maximum value of 
$P_k$ in the interval $0 < t_{\rm mod} < 200$ ms as a function 
of $\Omega$ (in units of $\omega_\perp$). Solid squares on the 
horizontal axis correspond to twice the eigenfrequencies $\omega(k)$ 
of the Bogoliubov equations, with $k=2\pi m /L$ and $m$ integer.}
\label{fig:visibility}
\end{figure}

At resonance the $\pm k$ components of the Fourier transform 
of the order parameter along the torus, $\tilde{\psi}(k,r_\perp)$, 
grow exponentially during the modulation. Since the GP equation 
(\ref{eq:GP}) contains a nonlinear mean-field term, the growth 
is limited by the energy transfer to 
nonresonant modes due to mode-mixing processes. Thus 
the $\pm k$ components eventually saturates around a maximum 
value for long times. An example is given in the upper panel 
of Fig.~\ref{fig:visibility}, where we show the quantity 
$P_k = 2 \pi \int dr_\perp \ r_\perp |\tilde{\psi}(k,r_\perp)|^2$
as a function of $t_{\rm mod}$ for $\Omega=0.6 \omega_\perp$. 
In the lower panel we plot the maximum value achieved by $P_k$ 
in case of a modulation with $t_{\rm mod}$ up to $200$ ms, as a 
function of $\Omega$. The position of each peak coincides with 
twice the frequency of the corresponding $m$-phonon, as 
calculated by solving the Bogoliubov equations. The figure 
suggests the possibility to use the parametric amplification 
for spectroscopy: Each time a density pattern is observed, the 
wavevector $k$ is simply obtained by counting the number 
of oscillations in the torus, while the frequency 
$\omega(k)$ is just $\Omega/2$ \cite{note-breathing}.

The standing density wave is phase-locked with the external 
modulation: the amplitude of density oscillations is maximum when 
the width of the condensate is either maximum or minimum, and 
vanishes when the width is the same of the unperturbed condensate. 
This also causes the fast oscillation of the quantity $P_k(t)$ in 
Fig.~\ref{fig:visibility}. The plots in Fig.~\ref{fig:exp-dens}(a)
and (d) correspond to a maximum and a minimum of $P_k(t)$, 
respectively; the density fluctuations in Fig.~\ref{fig:exp-dens}(d)
are random and small. The standing density wave is accompanied 
by a pattern in the velocity distribution. This is easily seen
by plotting the tangential current $J \equiv (i \hbar/2 M)
[\psi (\psi^*)^\prime - (\psi)^\prime \psi^*]$, as we did in
Fig.~\ref{fig:exp-dens}(b) and (e): the velocity distribution
exhibits small random fluctuations in (b) and a clean 
periodic pattern in (e). The density and velocity 
distributions oscillate exactly out-of-phase in time.

The occurrence of a pattern in the velocity field 
has spectacular consequences in the expansion after the 
release from the trap. In fact, the presence of $\pm k$-phonons 
gives rise to interference fringes of atoms expanding in 
preferred directions, similar to those observed in the expansion 
of an elongated condensate with Bragg excited phonons 
\cite{davidson}. In toroidal geometry these fringes produce 
a flower-like structure with $m$ ``petals" reflecting 
the periodicity of the initial velocities. An example is 
given in Fig.~\ref{fig:exp-dens}(c) and (f) where we 
show what happens to the condensates of Fig.~\ref{fig:exp-dens}(a) 
and (d), respectively, when the trap is switched off and the 
gas is imaged after $7$ ms of free expansion. The figure shows 
that, by starting the expansion when the pattern has maximum 
amplitude in velocity field, one obtains very clean interference 
fringes. Since these ``petals" are much more visible than the 
in-trap density oscillations, the expansion significantly 
enhances the sensitivity of the spectroscopic measurement.

\begin{figure}
\centerline{\includegraphics[height=0.5\columnwidth,clip=,angle=-90]{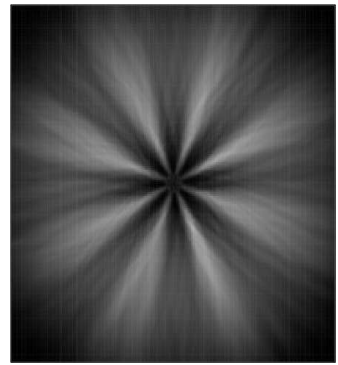}}
\caption{Density of the expanded condensate as in 
Fig.~\protect\ref{fig:exp-dens}(f). The expansion starts
again from the condensate in Fig.~\protect\ref{fig:exp-dens}(d), 
but here the mean-field interaction is included in the 
simulation of the expansion, while 
in Fig.~\protect\ref{fig:exp-dens}(f) it is neglected.  
The expansion time is $t_{\rm exp}
=1.8$ ms. The box is $45~\mu$m per side.}
\label{fig:exp-p-mf}
\end{figure}

The dynamics in Fig.~\ref{fig:exp-dens}(c) and (f) is 
simulated by solving Eq.~(\ref{eq:GP}) without the external 
potential and neglecting the mean-field interaction. From the 
experimental viewpoint, this situation corresponds to either  
tuning the scattering length $a$ to zero by means of a Feshbach 
resonance just before the expansion or starting from a torus 
strongly squeezed in the $z$-direction. In the latter case, due to 
the fast expansion along $z$, the density decreases much faster than 
the typical timescale for the transverse expansion, thus suppressing 
mean-field effects in the radial motion.  However 
the choice to neglect the mean-field interaction 
is not necessary for obtaining a distinct flower-like structure. 
This can be seen in Fig.~\ref{fig:exp-p-mf}  where we plot 
the density after free expansion including the mean-field interaction
for a condensate prepared as in Fig.~\ref{fig:exp-dens}(d); the 
petals are still visible also in this case. The main 
difference with Fig.~\ref{fig:exp-dens}(f) is the 
timescale of the expansion, which is faster in the presence of the 
repulsive mean-field interaction. A rough estimate of the ratio of 
the two timescales, which holds at short times, is easily obtained 
by comparing the expansion of a cylindrical condensate in these 
two cases: i) the scaling behavior of the full GP equation \cite{rmp} 
and ii) the dispersion of a noninteracting wavepacket with the same 
initial shape. Using the parameters of our condensate, the interacting 
gas turns out to expand faster by a factor $\gamma\simeq 3.3$. 
In a toroidal geometry this factor is expected to be slightly larger.
The ratio between the two expansion times in Fig.~\ref{fig:exp-dens} 
and Fig.~\ref{fig:exp-p-mf} has been taken to be $\simeq 4$ and 
the size of the expanding condensate and of the petals is indeed 
similar.

\begin{figure}
\centerline{\includegraphics[height=\columnwidth,clip=,angle=-90]{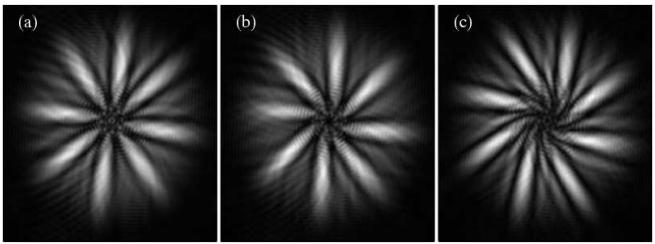}}
\caption{Interference patterns in the expansion of a parametrically 
excited condensate in the presence of quantized circulation ($\kappa=1,
5,10$ from $(a)$ to $(c)$). The superfluid rotation causes a misalignment 
of the fringes proportional to $\kappa$. In all plots $t_{\rm exp}=7$ ms 
and $t_{\rm mod}=180.8$ ms. Each box is $55~\mu$m per side.}
\label{fig:exp-qc}
\end{figure}

The process of pattern formation is significantly affected 
by the presence of quantized circulation. If the 
condensate is initially rotating with angular momentum $L_z
=\kappa\hbar$ per particle, where $\kappa$ is the quantum of 
circulation, then the external modulation produces a
pattern which rotates along the torus at the same angular 
velocity of the condensate as a result of the frequency shift 
of counter-rotating phonons \cite{zambelli}. More strikingly, the 
interference structure observed in the free expansion exhibits 
a significant misalignment of opposite petals, proportional 
to $\kappa$. This is evident in Fig.~\ref{fig:exp-qc} where we 
show the same expanded condensate but for different values of 
$\kappa$. The axis of each petal can be approximated with a 
rectilinear segment tangent to a circle of radius $R_0$, so 
that $2R_0$ is the distance between the axis of opposite 
petals. The radius $R_0$ is of the order of the size of a 
$\kappa$-vortex core in the expanding condensate. Its time 
evolution is well reproduced by the law $R_0(t) = R \kappa 
\alpha + \beta \kappa t/{R}$. The second term, with $\beta$ of 
order $1$, is consistent with the asymptotic expression for 
the vortex profile that can be derived analytically in the case 
of free expansion as discussed in \cite{cozzini}. The quantity 
$\kappa \alpha$ in the first term is of the order of the ratio 
between the rotational velocity of the condensate in the torus
and the velocity at which the condensate expands radially; this
ratio is $\kappa \alpha \sim (\hbar \kappa /MR)/(\omega_\perp 
r_0/\gamma)$. Fig.~\ref{fig:exp-qc} shows that 
the parametric amplification process can be used as 
a sensitive probe of rotations in the torus \cite{nota}, 
which works down to a few quanta of circulation and is 
complementary to the techniques discussed in \cite{cozzini}.

All results shown above have been obtained assuming the 
condensate to be at $T=0$. In this case, the fluctuations which 
are parametrically amplified are those due to the small quantum 
depletion of the condensate.  However the dynamics of the 
amplification is greatly independent of the type and origin of 
the initial fluctuations. What really matters is the amount of 
effective seed (i.e., a specific quadrature of counter-propagating 
Bogoliubov excitations \cite{tozzo}), which determines the time 
$t_{\rm mod}$ needed to obtain a visible density pattern. 
In an experiment the imperfections in loading the condensate in 
the trap and thermal fluctuations can also contribute to the 
seed. In particular, one expects that thermal fluctuations 
become dominant for temperature $T \gg \hbar \Omega/k_B$. We have 
performed simulations in this regime finding that the time
$t_{\rm mod}$ needed to achieve the maximum pattern visibility, defined
as in Fig.~\ref{fig:visibility}, scales approximately as $(1/\Omega) 
ln(\Omega/T)$. This is consistent with the fact that the growth
rate and the amount of seed are expected to be proportional 
to $\Omega$ and $T/\Omega$, respectively \cite{tozzo}. For the
parameters of Fig.~\ref{fig:visibility} and for $T \gg \hbar 
\Omega/k_B$ a factor $2$ in temperature implies a change of about 
$10$ ms in the timescale of the amplification. This suggests that the 
visibility of the pattern as a function of $\Omega$ could provide a
way to measure $T$ in a regime of temperatures where standard 
thermometric methods are not applicable.

In conclusion, we have shown that a periodic modulation of the 
confining potential of a toroidal condensate induces a spontaneous 
pattern formation through the parametric amplification of 
counter-rotating Bogoliubov excitations. This can be viewed as
a quantum version of Faraday's instability for classical fluids 
in annular resonators. The occurrence of this pattern in both 
density and velocity distributions provides a tool for measuring 
fundamental properties of the condensate, such as the excitation 
spectrum, the amount of thermal and/or quantum fluctuations and the 
presence of quantized circulation and persistent currents.


\begin{thebibliography}{}
\bibitem{note2D} We checked the accuracy of 
this approximation in the case of a 2D annular condensate, 
finding that the behavior of parametric resonances and 
pattern formation is scarcely affected by the inclusion of 
centrifugal terms.
\bibitem{nota}
We have verified that the misalignement has a similar behavior
and is still detectable also including interations during the 
expansion.
\end{thebibliography}
\end{document}